# LAYERWIDTH: Analysis of a New Metric for Directed Acyclic Graphs

**Mark Hopkins**
Department of Computer Science
University of California, Los Angeles
Los Angeles, CA 90095
mhopkins@cs.ucla.edu

## Abstract

We analyze a new property of directed acyclic graphs (DAGs), called layerwidth, arising from a useful class of DAGs proposed by Eiter and Lukasiewicz for tractable causal reasoning. First, we establish that the complexity of deciding whether a given graph has a bounded layerwidth is NP-complete. Then we proceed to prove key properties of layerwidth that are helpful in efficiently computing the optimal layerwidth. Finally, we compare this new DAG property to two other important DAG properties: treewidth and bandwidth.

## 1 INTRODUCTION

Halpern and Pearl [4, 5] have recently proposed a set of general-purpose definitions for cause and explanation. These definitions are embedded in the language of recursive structural models, the structure of which can be represented using a directed acyclic graph (DAG). In [1], Eiter and Lukasiewicz explored classes of DAGs for which Halpern and Pearl's definitions could be computed in polynomial time. In that work, they define what we will refer to as a *layer decomposition* of a DAG. They show that causes and explanations can be identified tractably in DAGs for which we have a layer decomposition of bounded width (given certain constraints on the query variables). We will formally define these concepts in the next section.

Their work leaves several questions open. Is it possible to compute the optimal layer decomposition (i.e. the layer decomposition of lowest width) of a given DAG in polynomial time? If not, how should such a decomposition be computed? Moreover, what is the relationship of the width of the optimal layer decomposition of a DAG to other popular graph metrics, such as treewidth?

In this paper, we strive to resolve these questions. We will begin by briefly reviewing the Halpern and Pearl definition of cause and discussing the tractable cases identified by Eiter and Lukasiewicz in [1]. Then we will formally define a DAG property called *layerwidth*, which is simply the width of the optimal layer decomposition of a DAG, and show that this concept is well-defined. We follow this by proving that the problem of computing the layerwidth of a DAG (and hence the problem of computing the optimal layer decomposition) is NP-complete. Given this intractability result, we provide a depth-first branch-and-bound algorithm for computing the optimal decomposition. This algorithm has the advantage of being an anytime algorithm, and hence can also be used as a heuristic if interrupted. Finally, we discuss the relationship of layerwidth to two other DAG properties, treewidth and bandwidth. In the interests of space, some proofs are abridged or omitted. These can be found in the full version of the paper [6].

## 2 STRUCTURAL CAUSAL MODELS AND LAYER DECOMPOSITIONS

Halpern and Pearl [4, 5] propose their definitions within the framework of structural causal models. Essentially, structural models are a system of equations over a set of random variables. We can divide the variables into two sets: endogenous (each of which have exactly one structural equation that determines their value) and exogenous (whose values are determined by factors outside the model, and thus have no corresponding equation).

Formally, a *structural causal model* (or *causal model*) is a triple $(\mathbf{U}, \mathbf{V}, \mathbf{F})$, in which $\mathbf{U}$ is a finite set of exogenous random variables, $\mathbf{V}$ is a finite set of endogenous random variables (disjoint from $\mathbf{U}$), and $\mathbf{F} = \{F_X | X \in \mathbf{V}\}$ where $F_X$ is a function $Dom(\mathbf{R}) \to Dom(X)$ that assigns a value to $X$ for each setting of the remaining variables in the model $\mathbf{R} = \mathbf{U} \cup \mathbf{V} \backslash \{X\}$. For each $X$, we can define $\mathbf{PA}_X$, the *parent set* of $X$, to be the set of variables in $\mathbf{R}$ that can affect the value of $X$ (i.e. are non-trivial in $F_X$). We also assume that the domains of the random variables are finite.

Causal models can be depicted as a *causal diagram*, a directed graph whose nodes correspond to the variables in $U \cup V$ with an edge from $Y$ to $X \in V$ iff $Y \in \mathbf{PA}_X$. We are specifically interested in *recursive causal models*, which are causal models whose causal diagram is acyclic.

Eiter and Lukasiewicz [1] have investigated classes of causal diagrams for which many of the causal queries proposed by [4] can be answered in polynomial time. These queries include actual cause, computing all actual causes, explanation, partial explanation, and $\alpha$-partial explanation. The details of these definitions are not directly relevant to this paper. This paper concerns itself with the classes of causal diagrams (directed acyclic graphs) for which these queries can be answered polynomially.

All of the tractable classes identified by Eiter and Lukasiewicz are subsumed by a class of directed acyclic graph that they refer to as *decomposable*. To understand this class of DAG, we need to define the concept of a *layer decomposition* of a DAG. The intuition behind a layer decomposition of a directed acyclic graph is to decompose the DAG into a chain of directed acyclic subgraphs that connect to one another through an independent set of interface variables. Formally, a *layer decomposition* of a DAG $G = (V, A)$ is a list $((T^0, S^0), ..., (T^k, S^k))$ of pairs $(T^i, S^i)$ of subsets of $V$ such that the following conditions hold [1]:

D1. $(T^0, ..., T^k)$ is an ordered partition of $V$.

D2. $S^0 \subseteq T^0, ..., S^k \subseteq T^k$.

D3. For every $i \in \{0, ..., k-1\}$, no two variables $A \in T^0 \cup ... \cup T^{i-1} \cup T^i \backslash S^i$ and $B \in T^{i+1} \cup ... \cup T^k$ are connected by an arrow in $G$.

D4. For every $i \in \{1, ..., k\}$, every child of a variable in $S^i$ in $G$ belongs to $(T^i \backslash S^i) \cup S^{i-1}$. Every child of a variable in $S^0$ belongs to $(T^0 \backslash S^0)$.

D5. For every $i \in \{0, ..., k-1\}$, every parent of a variable in $S^i$ in $G$ belongs to $T^{i+1}$. There are no parents of any variable $A \in S^k$.

Figure 1 depicts a DAG and three layer decompositions of it. We will refer to each $T_i$ as a *block* of the layer decomposition. For example, we can refer to $T_4$ as the 4th block of the decomposition. We will refer to each $S_i$ as the *interface* of the corresponding block. Occasionally we will informally refer to $T^0$ as the "rightmost" block of the layer decomposition, and $T^k$ as the "leftmost" block, stemming from our convention of graphically depicting layer decompositions (for instance, in Figure 1).

The definition is identical to the decomposition presented by Eiter and Lukasiewicz in [1], except that we do not constrain the placement of any variables of the DAG. Eiter and Lukasiewicz require that certain root variables are constrained to be in the leftmost ($k$th) block, while another subset of variables are constrained to be in the rightmost (0th) block. We will address the impact of such constraints later. For now, we consider the more basic problem.

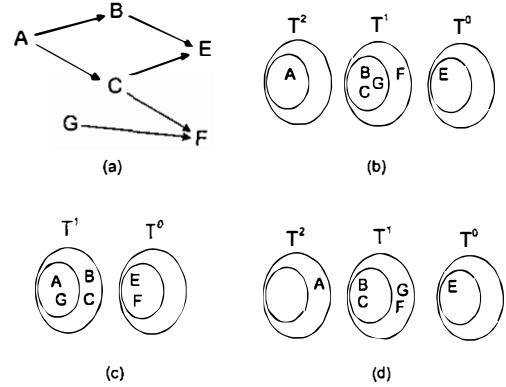

Figure 1: A DAG $G$ and three layer decompositions of $G$ of width 4. The subellipse inside $T^i$ represents $S^i$.

We define the *width* of a layer decomposition as the lowest integer $w$ such that $|T^i| \leq w$ for every $i \in \{1, ..., k\}$.

Notice that every layer decomposition has width at least 1, and that every DAG $G = (V, A)$ has the trivial decomposition $((\emptyset, V))$. Hence it is well-defined to talk about the lowest width layer decomposition that exists for a particular DAG $G$. We refer to the width of such a decomposition as the *layerwidth* of $G$.

## 3 COMPLEXITY RESULTS

We now define the following problem:

**LAYERWIDTH**
INSTANCE: Directed acyclic graph $G$, positive integer $k$.
QUESTION: Does there exist a layer decomposition of $G$ of width $\leq k$?

We will show that this problem is NP-complete. It is clear that this problem is in NP, since as a certificate we can simply present a (polynomial size) layer decomposition of width $k$ or less, which can be "guessed" and verified in polynomial time by a nondeterministic Turing machine. Thus our main task is to prove that the problem is NP-hard.

We will prove this via a reduction from 3-PARTITION, which is defined as follows [3]:

**3-PARTITION**
INSTANCE: Set $A$ of $3m$ elements, a bound $D \in Z^+$, and a size $s(a) \in Z^+$ for each $a \in A$ such that $D/4 < s(a) < D/2$ and such that $\Sigma_{a \in A} s(a) = mD$.

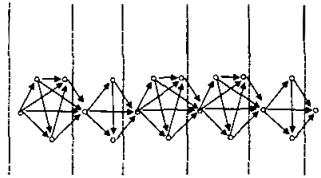

Figure 2: A chain of directed cliques, subdivided into segments.

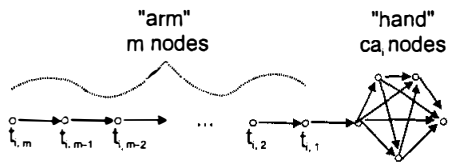

Figure 3: A tentacle corresponding to 3-partition element $a_i$.

QUESTION: Can $A$ be partitioned into $m$ disjoint sets $A_1, A_2, ..., A_m$ such that, for every $i \in \{1, ..., m\}$, $\Sigma_{a \in A_i} s(a) = D$?

Notice that because of the constraint on $s(a)$, each set must contain exactly 3 elements of $A$. Thus the goal of this problem is to see if it is possible to partition the set of $3m$ elements into $m$ 3-element sets that each add up to $D$.

For example, say that we have the set of elements $\{6, 6, 6, 6, 7, 8, 9, 10, 11\}$, and bound $D = 23$. A valid 3-partition exists for this set of elements, namely $A_1 = \{6, 8, 9\}, A_2 = \{6, 6, 11\}, A_3 = \{6, 7, 10\}$. Note that $6 + 8 + 9 = 23, 6 + 6 + 11 = 23$, and $6 + 7 + 10 = 23$.

Our reduction is inspired by the reduction proof of [2], which shows the NP-hardness of computing the minimum bandwidth of a tree of degree 3. Their construction uses a special kind of tree which they dub "siphonophoric," due to its similarities with pelagic hydrozoa of the order Siphonophora. Our construction bears less of a resemblance to aquatic life, however we will borrow liberally from their terminology, when appropriate.

We will need the notion of *a chain of directed cliques*. To construct a *directed clique* over a set of nodes $W = \{w_0, ..., w_k\}$, we add arrows such that there is an arrow from $w_i$ to $w_j$ if (and only if) $i > j$. The *sink* of this clique is $w_0$ and the *source* of this clique is $w_k$. We will call the set $W \setminus \{w_0\}$ the *segment* of $W$ and define the source of segment $W \setminus \{w_0\}$ to be the source of the clique $W$. For segment $W$, we will use the notation $W^{src}$ to denote the singleton set containing the source of $W$.

A *chain of directed cliques* is a minimal DAG $G$ over sets of nodes $(\{w_0\}, W_0, W_1, ..., W_l)$ such that $W_0 \cup \{w_0\}$ is a directed clique with sink $w_0$, $W_1 \cup W_0^{src}$ is a directed clique with sink $W_0^{src}$, ..., $W_l \cup W_{l-1}^{src}$ is a directed clique with sink $W_{l-1}^{src}$. By minimal, we mean that $G$ does not contain any edges except those required to satisfy the conditions stipulated above, e.g., $G$ does not contain any arrows from $W_5$ to $W_3$. We will refer to each $W_i$ as the *segments* of the chain and call $w_0$ the *tip* of the chain. We show a chain of directed cliques in Figure 2.

Given an instance of 3-PARTITION, we will now construct a DAG $G$ such that the layerwidth of $G$ is $k$ (or less) if and only if the instance has a satisfying 3-partition (where $k$ is some value that will be fixed shortly). We begin by constructing the so-called *body* of our graph.

The *body* of the graph will consist of a chain of directed cliques over $(\{p_0\}, P, B_1, B_2, ..., B_m, H)$ such that:

- $|P| = k$.
- For every $i \in \{1, 2, ..., m\}$, $|B_i| = k - (6i - 3) - cD$.
- $|H| = k$.

where $c = 3m^2 + 9m$ and $k = 2(6m - 3 + cD) + 1$. The specific values of $c$ and $k$ are not important to worry about now, except to show that we can construct the body of the graph in polynomial time. For this, we must observe that 3-PARTITION is "strongly" NP-complete [3], which for our purposes means that 3-PARTITION remains NP-complete, even when we restrict our focus to instances such that $D$ is bounded above by a (suitably large) polynomial function of $m$. Thus, we need only show that the size of the graph we are constructing is polynomial in $m$ and $D$.

The body has $1 + 2k + \Sigma_{i=1}^{m}(k - (6i - 3) - cD) = 9m^2D + 54mD + 36m - 3m^3D - 3m^2$ vertices, thus it can be constructed in time polynomial in $m$ and $D$. We will refer to segment $H$ as the *head* of the body, segments $B_i$ as the *spine* of the body, and segment $P$ as the *tail* of the body.

Now for each $a_i \in A$, we construct a *tentacle* of the graph. A *tentacle* will consist of a chain of $m$ nodes attached to the source of a directed clique of $ca_i$ nodes, shown in Figure 3. We refer to the directed clique at the end of each tentacle as the *hand* of the tentacle. We refer to the chain of $m$ nodes as the *arm* of the tentacle. For the tentacle corresponding to 3-partition element $a_i$, we refer to the node of the arm that is $j^{th}$ closest to the hand as $t_{i,j}$. For example, the arm node closest to the hand is $t_{i,1}$ and the arm node closest to the head is $t_{i,m}$. Each tentacle is attached to one of the nodes (it does not matter which) in the head of the body, i.e. $t_{i,m}$ is the child of an arbitrary node of the head.

Notice that the tentacles contain $3m * m = 3m^2$ arm nodes, and $cmD = 3m^3D + 9m^2D$ hand nodes, thus these can also be constructed in time polynomial in $m$ and $D$.

This completes the description of the construction of the DAG $G$ corresponding to an instance of 3-PARTITION. Observe that for a fixed $m$ and $D$, the only difference between graphs corresponding to different instances is the size of the hands of the tentacles. Intuitively, we are trying to fit exactly three hands into each block of the layer decomposition containing a spine segment. It can be shown that this is possible if (and only if) there exists a valid 3-partition for the instance.

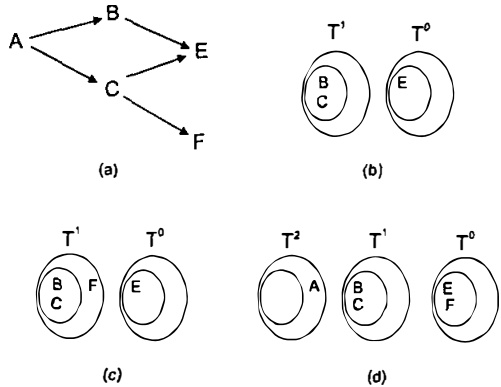

Figure 4: A DAG G (a) and three PLDs of G (b,c,d).

The proof proceeds roughly as follows. Suppose that there exists some layer decomposition $D$ of $G$ with width $k$. First observe that the body is a chain of directed cliques, and that each segment of the body must appear in a single block of $D$. In other words, no body segment can span two blocks. Furthermore, the head segment must appear in the leftmost block and the tail segment must appear in the rightmost block of D (actually, the second to rightmost – the tip appears in the rightmost block). Moreover, each spine segment must appear in its own block (in between the head and tail blocks) since any two spine segments contain more than $k$ variables.

Since the head and tail blocks each contain $k$ variables, the tentacles must squeeze into the available space in the blocks occupied by the spine segments. In fact, there is just enough space in these blocks to accommodate the variables of the tentacles. The proof proceeds to show that the tentacles fit in these blocks if (and only if) a valid 3-partition exists for the instance of 3-PARTITION that the DAG corresponds to. In this case, we can fit three hands into each block containing a spine segment. We must be careful about how the arms fit in – the choice of $c$ is chosen such that the proof works. The complete proof is available in the full version of the paper [6].

**Theorem 1** *Suppose that we have an instance of 3-PARTITION, and that G is the DAG corresponding to the construction outlined above. There exists a valid 3-partition for this instance if (and only if) the layerwidth of G is at most k.*

Given what we have established, the following theorem is immediate:

**Theorem 2** *LAYERWIDTH is NP-complete.*

In the definition of layer decompositions proposed in [1], there is an additional constraint on the definition to allow causes to be tractably identified. Namely, the "cause" variables of the causal network (DAG) must be placed in the interface of the leftmost block of the layer decomposition, and the "effect" variables must be placed in the rightmost block of the layer decomposition. With the above result in hand, it is a straightforward exercise to prove that the problem of finding the optimal layer decomposition of a DAG subject to such constraints is also NP-complete. The details can be found in the full version of the paper [6].

## 4 COMPUTATION

In the previous sections, we have established the intractability of finding the optimal layer decomposition for a given DAG. In this section, we consider how we can compute such a decomposition as efficiently as possible. To this end, we propose a depth-first branch-and-bound algorithm. In choosing this approach, we gain the advantage of interruptability, i.e. the computation can be stopped at any point and will return the best result it has found thus far. Hence it can also be used as a heuristic algorithm if run-time is constrained.

We need to first establish a few preliminary definitions. First, we define a *partial layer decomposition (PLD)* of a DAG $G = (V, A)$. This is simply a layer decomposition of $G[W]$, where $G[W]$ is the subgraph of $G$ over a subset of variables $W \subseteq V$ (consisting of $W$ and the arrows of $G$ that both originate from a node in $W$ and terminate at a node in $W$). We will refer to the set $W$ as $Vars(D)$, where $D$ denotes the PLD. Figure 4(b) shows a PLD $D$ of a DAG $G$ such that $Vars(D) = \{B, C, E\}$. Since the PLD is a layer decomposition of $G[\{B, C, E\}]$, we can further define the $width$ of a PLD to be the width of this layer decomposition. The width of the PLD in Figure 4(b) is 2.

Second, we define a *sub-PLD* $D'$ of a PLD $D$ of DAG $G = (V, A)$. Simply put, a sub-PLD $D'$ of a PLD $D$ is a layer decomposition over a subset of the variables in $D$ that maintains the relative positions of these variables. In formal terms, let $D = ((T_1^0, S_1^0), (T_1^1, S_1^1), ..., (T_1^k, S_1^k))$ be a PLD of $G$. Let $D' = ((T_2^0, S_2^0), (T_2^1, S_2^1), ..., (T_2^j, S_2^j))$ be a PLD of $G$ such that $Vars(D') \subseteq Vars(D)$. Then $D'$ is a *sub-PLD* of PLD $D$ iff there exists some non-negative integer $m$ such that for all $i \in \{0, 1, ..., j\}$, we have that $T_2^i \subseteq T_1^{i+m}$ and $S_2^i \subseteq S_1^{i+m}$. This definition is a bit hard to parse, but the intuition behind it is quite straightforward. Figure 4(b) is a sub-PLD of Figure 4(c) and Figure 4(d) since the relative positions of $B$, $C$, and $E$ are maintained, but notice that Figure 4(c) is not a sub-PLD of Figure 4(d).

Third, given DAG $G = (V, A)$ and a PLD $D$ of $G$, an *insertion* of variable $X \in (V \setminus Vars(D))$ into $D$ is a new PLD $D'$ of $G$ such that (a) $Vars(D') = Vars(D) \cup \{X\}$ and (b) $D$ is a sub-PLD of $D'$. Furthermore, to *insert* vari-able $X$ into PLD $D$ is to produce an insertion of $X$ into $D$. For example, Figure 4(c) is an insertion of variable $F$ into Figure 4(b).

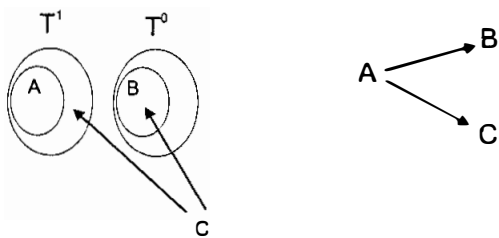

Figure 5: Since A is the parent of C, C's insertion into any PLD containing A is constrained to two unique positions.

Finally, we need the concept of a *boundary variable*. Given a DAG $G = (V, A)$ and a subset $W \subseteq V$, we define a *boundary variable* of $W$ as any variable $X \in V \setminus W$ such that some parent or child of $X$ in DAG $G$ is a member of $W$. For example, for the DAG in Figure 4(a), the boundary variables of $\{B, E\}$ are $A$ and $C$ (but not $F$).

Now to establish our search space, we need to prove a theorem. This theorem essentially allows our search space to be a binary search tree.

**Theorem 3** *Let $D$ be a PLD of DAG $G = (V, A)$. Let $X \in V$ be a boundary variable of $Vars(D)$. Then there exist at most two unique insertions of $X$ into $D$.*

We omit a formal proof here in favor of motivating intuition. Consider Figure 5, which shows a PLD of the DAG shown in Figure 4(a) over $\{A, B\}$. There are only two valid ways to insert $C$ into this PLD. If we attempt to have $C$ occupy any other position of the PLD, we violate condition D4 of the definition. Similarly, if $A$ were elsewhere, or if $A$ were $C$'s child in the DAG, we can prove through a detailed case analysis that there are at most two insertions of $C$ into the PLD.

This means that we can represent all possible layer decompositions of DAG $G$ as a binary search tree. Suppose that an internal node of the search tree corresponds to some PLD over a proper subset of variables $W$ of $G$. At the subsequent level, we choose a boundary variable of $W$ and produce all possible PLDs that result from inserting this variable into the PLD. From the theorem, there are only two of these. Clearly as long as $G$ is connected and $W$ is non-empty, there will always be some boundary variable to choose. But what about the base case, when $W$ is empty? The following theorem gives us our starting point.

**Theorem 4** *Let $G = (V, A)$ be a directed, acyclic graph. Let $X \in V$. Then there exist exactly two unique PLDs $D$ of $G$ such that $Vars(D) = \{X\}$.*

**Proof** Simply put, the two PLDs take the form $((T_0, S_0))$.

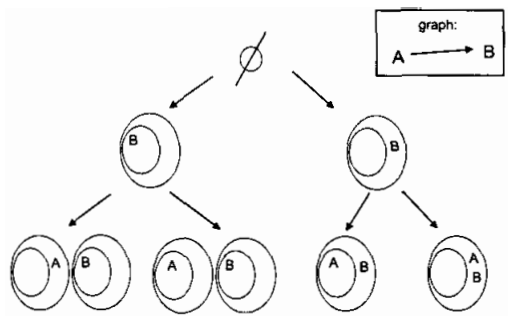

Figure 6: A complete search tree for a simple chain graph over two variables.

In one, $T_0 = \{X\}$ and $S_0 = \{X\}$. In the other, $T_0 = \{X\}$ and $S_0 = \emptyset$. There are no other PLDs over a single variable. ∎

Hence we have established our search space. At the root, we begin with the trivial PLD over the empty set, and at each subsequent level of the search tree, we insert some boundary variable into the PLDs that we have generated at the previous level of the tree. Figure 6 shows a complete search space for a simple chain graph over two variables.

There is no reason that we need to insert nodes to our PLD in a fixed order down every path of our search tree. Instead, at each node of our search tree we can dynamically choose to insert any node of the graph that has a parent or child that has already been inserted to the PLD (recall that this condition restricts the number of possible insertions to at most two). This strategy is advantageous because we can first add any nodes for which there is only one possible insertion, given the current PLD. Furthermore, if any nodes exist for which there is no possible insertion, we can immediately return $nil$ (meaning that no layer decomposition subject to the given constraints exists). We will refer to this process as *resolution*. These observations give rise to the following basic algorithm:

Let $G = (V, A)$ be a directed, acyclic graph. Then the call $BasicLD(G, \emptyset)$ returns an optimal layer decomposition of $G$, where algorithm $BasicLD$ is defined as follows:

Algorithm $BasicLD$( DAG $G$, PLD $D$ ):

1. Let $D = Resolution(G, D)$.

2. If $Vars(D) = V$, then return $D$.

3. If $D = nil$ then return $nil$.

4. If $Vars(D) = \emptyset$ then let $X$ be any node of $G$; otherwise let $X \in V$ be any boundary variable of $Vars(D)$.

5. For every insertion $D_i$ of $X$ into $D$: Let $F_i = BasicLD(G, D_i)$.

6. If all $F_i = nil$, then return $nil$. Otherwise, return the layer decomposition $F_i$ of minimum width.

For now, we defer a precise consideration of the function $Resolution(G, D)$, except to say that it returns nil if there is some variable of $G$ that cannot be inserted into $D$ and otherwise recursively places all variables of $G$ for which there is only one possible insertion until all variables of $G$ that are not in $Vars(D)$ have at least two possible insertions.

**Theorem 5** $BasicLD(G, \emptyset)$ returns an optimal layer decomposition of $G$.

**Proof sketch** Consider the search tree of $BasicLD(G, \emptyset)$. Suppose that $N$ is a node of this search tree corresponding to the call $BasicLD(G, D)$. Notice that if $N$ is at level $k$ of the search tree, then $D$ is a PLD of $G$, by an easy inductive argument. Define $PLD(N) = D$. Notice further that for any leaf node $N$ of the search tree, $PLD(N)$ corresponds to a layer decomposition of $G$.

Clearly then, $BasicLD(G, \emptyset)$ returns the lowest-width layer decomposition of $G$, among the layer decompositions represented by the leaves of the search tree. Thus to prove that it returns the optimal layer decomposition of $G$, we need only show that every layer decomposition of $G$ is represented by some leaf of the search tree.

We can prove this by induction. Fix any layer decomposition $D$ of $G$. We want to show that if there exists some node $N$ such that $PLD(N)$ is a sub-PLD of $D$, then either $PLD(N) = D$ (in which case $N$ is a leaf node), or $N$ has a child $N'$ such that $PLD(N')$ is also a sub-PLD of $D$. Notice that for root $R$ of the search tree, $PLD(R) = \emptyset$, which is a sub-PLD of every layer decomposition of $G$. Thus if we can prove the above statement, then we will have proven that there exists some leaf $N$ of the search tree such that $PLD(N) = D$. The details of this induction are available in the full version of the paper. [6] ∎

We claim that the worst-case time complexity is $O(2^n * poly(n))$, where $n$ is the number of nodes of $V$ and $poly(n)$ is a polynomial functional of $n$. Since we have already established that the search tree has $O(2^n)$ nodes (from Theorem 3), we need only show that a polynomial amount of work is done at each node. This is relatively trivial, since steps 2, 3, and 4 can clearly be performed in polynomial time, while step 6 requires us to be able to compute the width of a given layer decomposition, which can easily be shown to be polynomial. Step 5 requires us to generate all insertions of a boundary variable into a PLD. From Theorem 3, at most two such layer decompositions exist. They are also easy to generate, since we are essentially just adding a node to the existing layer decomposition. We will further assume that $Resolution(G, D)$ runs in polynomial-time, thus $BasicLD$ runs in time $O(2^n * poly(n))$.

Let us now turn our attention to the important resolution step. It is not hard to go through each of the boundary variables and assess which have zero or one possible insertion, given the constraints placed upon them by previously inserted parents and children. But is this all that we can do? It turns out that there exists a non-trivial class of graph nodes which we can automatically insert, even if there seems to be two possible insertions for the node.

**Theorem 6** Let $D$ be a sub-PLD of DAG $G$. Let $X$ be a root variable of $G$ such that $X$ is a boundary variable of $Vars(D)$. Define $w(D')$ to be the width of the optimal layer decomposition $D''$ such that $D'$ is a sub-PLD of $D''$. Then $w(D')$ is the same for every insertion $D'$ of $X$ into $D$.

**Theorem 7** Let $D$ be a sub-PLD of DAG $G = (V, A)$. Let $X \in V$ be a boundary variable of $Vars(D)$. If in $G$, any ancestor of $X$ is directly connected to any descendant of $X$, then there exists at most one insertion $D'$ of $X$ into $D$ such that $D'$ is a sub-PLD of a layer decomposition of $G$.

The upshot of these two theorems is that we do not have to branch on two special classes of DAG variables: root variables, and variables that have any ancestor directly connected to any descendant. This can in fact constitute a large proportion of the variables in a given DAG. Notice that both of these variable sets can be determined statically in polynomial time simply by looking at the structure of the DAG. Hence using a resolution function that utilizes these theorems means that BasicLD has running time $O(2^m * poly(n))$, where $m$ is the number of DAG variables that are neither roots nor have an ancestor directly connected to a descendant.

We have now developed a depth-first search algorithm whose goal is to find the leaf of minimum width in a tree of known depth. Hence, this algorithm is an ideal candidate to transform into a branch-and-bound algorithm. To do so, we need a cost function $g(N)$ for each internal node $N$ and a heuristic function $h(N)$ that is a lower-bound on the minimum-width layer decomposition that is a descendant of $N$. In this case, it is convenient to set $g(N) = 0$ for all internal nodes $N$ and simply focus on how to establish a tight lower bound on the lowest possible width it is possible to achieve, starting with the PLD represented by search tree node $N$ (which we denote $PLD(N)$).

Clearly for a given node $N$ of the search tree, $PLD(N)$ is a sub-PLD of every layer decomposition represented by a descendant leaf. Thus the width of $PLD(N)$ is a lower bound on the best width of any layer decomposition represented by a descendant in the search tree. Hence we could set $h(N)$ to the width of $PLD(N)$. This is conceptually simple and straightforward to compute. However, we can do better than this. Suppose that in $G$, there is a parent $X$ of some variable $Y \in Vars(PLD(N))$ such that $X \notin PLD(N)$. In other words, we have inserted $Y$ in the layer decomposition, but not its parent. What can we say about where its parent must be inserted? Although we cannot say for certain the specific location of $X$, we can say precisely which block $X$ must end up in (though it may or may not be in that block's interface). But all we need to compute the width of the resulting layer decomposition is to know which variables are in which block. Thus we can compute $h(N)$ to be the width of $PLD(N)$ once the uninserted parents of $Vars(N)$ are added to their corresponding blocks.

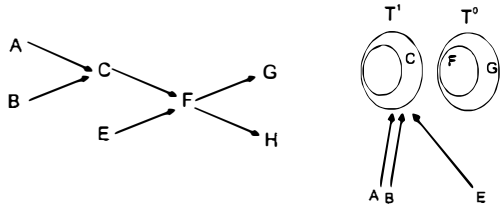

Figure 7: A DAG $G$ and a PLD of $G$ over $\{C, F, G\}$. Clearly, any layer decomposition that we can obtain by inserting $\{A, B, E, H\}$ into the PLD must have width at least 2, since this is the width of the PLD. However, the parent set of $\{C, F\}$ all must share a block with $C$, thus any layer decomposition that we can obtain by insertion must have width at least 4.

**Theorem 8** *Let $D$ be a sub-PLD of DAG $G$. Let $X$ be a variable of $G$ such that $X \notin Vars(D)$ but such that at least one child $Y$ of $X$ in DAG $G$ is a member of $Vars(D)$. Let $D' = ((T^0, S^0), ..., (T^k, S^k))$ be any layer decomposition of $G$ such that $D$ is a sub-PLD of $D'$. Then if $Y \in S^i$ for some $i \in \{0, ..., k-1\}$, then $X \in T^{i+1}$. Otherwise, if $Y \in (T^i \setminus S^i)$ for some $i \in \{0, ..., k\}$, then $X \in T^i$.*

**Proof** Suppose $Y \in S^i$ for some $i \in \{0, ..., k-1\}$. Then by D5, $X \in T^{i+1}$. Suppose $Y \in (T^i \setminus S^i)$ for some $i \in \{0, ..., k\}$. Then by D3 and D4, $X \in T^i$. Notice that $Y$ cannot be a member of $S^k$, otherwise D5 is necessarily violated. ∎

We show an example of the heuristic resulting from Theorem 8 in Figure 7. Using this heuristic to turn our existing algorithm into a depth-first branch-and-bound algorithm is simple. At every node $N$ such that $h(N)$ is greater than or equal to the best width found thus far in the computation, return nil and do not proceed to explore the subtree rooted at $N$. Since the heuristic is admissible, our algorithm maintains its optimality.

Hence in this section, we have developed a depth-first branch-and-bound algorithm for determining the optimal layer decomposition of a directed acyclic graph. This algorithm benefits from a number of important properties that we have proven about layer decompositions. The algorithm has the added advantage of being *anytime*. In other words, it finds a solution as soon as it hits a leaf, and from then on, giving the algorithm extra time simply makes the solution better, until the computation is interrupted or completed. Finally, it is easy to adapt this algorithm to the situation in which we have constraints on where certain variables must be placed in the final layer decomposition. To do so, we simply ignore any leaf representing a layer decomposition that does not comply with our constraints.

## 5 COMPARISON WITH OTHER DAG PROPERTIES

In this section, we compare the layerwidth of a DAG with two other important DAG properties: treewidth and bandwidth. Notice that both treewidth and bandwidth also have definitions for undirected graphs, but here we are concerned with directed, acyclic graphs. We will show that, in general, treewidth and layerwidth are non-comparable in the sense that neither dominates the other. For example, there exists a DAG whose treewidth exceeds its layerwidth, and there also exists a DAG whose layerwidth exceeds its treewidth. The same can be said of the relationship between bandwidth and layerwidth.

The *treewidth* of a DAG can be defined in a number of different ways. We will define it here in terms of elimination orders. Consider a DAG $G = (V, A)$. First, we must moralize the DAG, i.e. pairwise connect all parents of every node, then drop directionality from all edges of the graph. An *elimination order* of $G$ is simply any ordering of the variables in $V$. To *eliminate* a variable $X \in V$ from $G$, we pairwise connect all neighbors of $X$, then remove $X$ from the graph along with any incident edges. The *width* of an elimination order $\sigma$ is the maximal number of neighbors that any node has at its point of elimination, if we eliminate the nodes in the order perscribed by $\sigma$. The *treewidth* of a DAG $G$ is the lowest width among all elimination orders of $G$.

**Theorem 9** *If the layerwidth of a DAG $G$ is $w$, then the treewidth of $G$ is less than or equal to $2w - 1$. Furthermore, this bound is strict, i.e. for every $w$, there exists a DAG $G$ with layerwidth $w$ and treewidth $2w - 1$.*

**Proof sketch** Suppose that $D = ((T^0, S^0), ..., (T^k, S^k))$ is a layer decomposition of $G$ of width $w$. Let $\sigma$ be an elimination order of $D$ such that for $i \in \{0, ..., k-1\}$, all the variables in $T^i$ appear before all the variables in $T^{i+1}$ in $\sigma$. We want to show that the width of $\sigma$ is at most $2w - 1$. By induction, we can easily prove that at the point of elimination, any variable in $T^i$ can only be connected to variables in $T^i$ or $T^{i-1}$, which is a total of $2w - 1$ variables (not including itself).

To see that this bound is strict, consider the DAG $G = (V_1 \cup V_2 \cup \{X\}, A)$ where $V_1$ and $V_2$ are independent sets of $w$ variables each, every variable in $V_1$ is the parent of every variable in $V_2$, and every variable in $V_2$ is the parent of $X$. Clearly $D = ((T^0, S^0), (T^1, S^1), (T^2, S^2))$ where $T^0 = S^0 = \{X\}$, $T^1 = S^1 = V_2$, and $T^2 = S^2 = V_1$ is a layer decomposition of $G$ of width $w$. However the moral graph of $G$ contains a clique of size $2w$ over $V_1 \cup V_2$, hence the treewidth of $G$ is at least $2w - 1$. ∎

We actually cannot provide a bound in the opposite direction. In fact, there are graphs of treewidth 1 whose layerwidth is $|V|/2$. Namely, the rooted, directed tree of height 1 (with $|V| - 1$ leaves) has this property. Furthermore, there are graphs of treewidth 2 whose layerwidth is $|V| - 1$ (the worst possible layerwidth). Specifically, a chain of $|V|$ nodes where the root node is connected to the terminal node has this property.

Now we turn our attention to *bandwidth*. To define the bandwidth of a DAG, we will first review the concept of a topological order. A *topological order* $\phi$ of a DAG $G = (V, A)$ is an ordering of the variables in $V$ such that if $X \in V$ is a parent of $Y \in V$, then $X$ appears before $Y$ in $\phi$. We define the *width* of a topological order to be the maximum distance between a parent and its child in the order. For instance, for the DAG pictured in Figure 7, the width of topological order $A, B, E, C, F, G, H$ is 3, since $C$ is in position 4 and $A$ is in position 1. The bandwidth of a DAG $G$ is the lowest width among all topological orders of $G$.

**Theorem 10** *If the layerwidth of a DAG $G$ is $w$, then the bandwidth of $G$ is less than or equal to $2w - 1$. Furthermore, this bound is strict, i.e. for every $w$, there exists a DAG $G$ with layerwidth $w$ and bandwidth $2w - 1$.*

**Proof** Suppose that $D = ((T^0, S^0), ..., (T^k, S^k))$ is a layer decomposition of $G$ of width $w$. Let $\phi$ be an topological order of $D$ such that for $i \in \{1, ..., k\}$, all the variables in $T^i$ appear before all the variables in $T^{i-1}$ in $\sigma$. Variables in $T^0$ can only be connected to other variables in $T^0$. Moreover, for $i \in \{1, ..., k\}$, all the variables in $T^i$ can only be connected to the variables in $T^i \cup T^{i-1}$. Thus the furthest distance between a parent and a child in $\phi$ is $2w - 1$ (the number of variables in $T^i \cup T^{i-1}$, minus one).

To see that this bound is strict, consider DAG $G = (V_1 \cup V_2, A)$ where $V_1$ and $V_2$ are independent sets of $w$ variables each, and every variable in $V_1$ is the parent of every variable in $V_2$. Clearly $D = ((T^0, S^0), (T^1, S^1))$ where $T^0 = S^0 = V_2$, and $T^1 = S^1 = V_1$ is a layer decomposition of $G$ of width $w$. However in any topological order of $G$, some variable of $V_1$ must appear first, and some variable of $V_2$ must appear last. Thus every order has width $2w - 1$. ∎

Just as with treewidth, we cannot provide a bound in the opposite direction. There are graphs of bandwidth 2 whose layerwidth is $|V| - 1$ (the worst possible layerwidth).

## 6 DISCUSSION

In this paper, we have provided a detailed analysis of a DAG decomposition called a layer decomposition, recently proposed by Eiter and Lukasiewicz [1]. As we have mentioned, many intractable problems of causality and explanation in structural models have been found to be tractable for structural models whose DAG representation has a layer decomposition of bounded width [1].

Here, we have considered the problem from a broader perspective – as a general property of DAGs called layerwidth. As such, any intractable DAG problem can potentially benefit from the analysis presented here. This raises the question: the subset of DAGs of bounded layerwidth is an attractive subset for what kind of DAG problems (besides structural model-based causality)? It is hard to give specifics, but one possibility might be problems concerning dynamic Bayesian networks, whose structure lends itself to decomposition into layers. In any event, we have sought in this paper to establish layerwidth as a new metric for the toolbox of researchers designing DAG algorithms.

## Acknowledgments

This research was supported in part by grants from NSF, ONR, AFOSR, and the DoD MURI program.